\documentclass[12pt]{article}
\usepackage{paper2e}
\usepackage{mydefs2e}

\newcommand{\mm}{~\hbox{\rm mm}}

\begin{document}
\begin{titlepage}
\preprint{UMD-PP-00-057}

\title{A Critical Cosmological Constant\\\medskip
from Millimeter Extra Dimensions}


\author{Jiunn-Wei Chen,%
\footnote{\tt jwchen@physics.umd.edu}
\ 
Markus A. Luty,%
\footnote{\tt mluty@physics.umd.edu}
\ 
Eduardo Pont\'on%
\footnote{\tt eponton@wam.umd.edu}}

\address{Department of Physics\\
University of Maryland\\
College Park, Maryland 20742, USA}

\begin{abstract}
We consider `brane universe' scenarios with standard-model fields localized
on a 3-brane in 6 spacetime dimensions.
We show that if the spacetime is rotationally symmetric about the brane,
local quantities in the bulk are insensitive to the couplings on
the brane.
This potentially allows compactifications where the effective 4-dimensional
cosmological constant is independent of the couplings on the 3-brane.
We consider several possible singularity-free compactification mechanisms,
and find that they do not maintain this property.
We also find solutions with naked spacetime singularities, and we
speculate that new short-distance physics can become important near the
singularities and allow a compactification with the desired properties.
The picture that emerges is that standard-model loop
contributions to the effective 4-dimensional cosmological constant can be
cut off at distances shorter than the compactification scale.
At shorter distance scales, renormalization effects due to
standard-model fields renormalize the 3-brane tension, which changes a deficit
angle in the transverse space without affecting local quantities in the bulk.
For a compactification scale of order
$10^{-2}\mm$, this gives a standard-model contribution to the
cosmological constant in the range favored by cosmology.
\end{abstract}

\date{Revised June 23, 2000}

\end{titlepage}


\section{Introduction}
The cosmological constant problem is by far the most severe fine-tuning
problem in physics.
Despite many interesting proposals (see \Ref{Weinberg} for a
review) the problem still lacks a compelling solution.
Recently, there has been an interesting idea in the context of the
`brane universe' scenario that may solve part of the cosmological constant
problem, namely the sensitivity of the effective low-energy cosmological
constant to standard-model radiative corrections \cite{ADKS,KSS}.
In this proposal the standard-model fields are assumed to be localized
on a 3-brane in 5 spacetime dimensions.
It was argued in \Refs{ADKS,KSS} that special bulk-brane interactions can be
chosen so that solutions with unbroken 4-dimensional Poincar\'e invariance
exist independently of the standard model parameters and the 3-brane tension.
The solutions found in \Refs{ADKS,KSS} have a naked spacetime
singularity, whose resolution in a more fundamental theory of gravity
was argued to play an important role in this scenario.
On the other hand, the description of the cancellation mechanism of the
cosmological constant from a low-energy field theory perspective is
not evident.
Also, it is not clear whether these ideas can be extended to eliminate
the fine-tuning of bulk interactions \cite{Nilles}.

In this paper we describe a mechanism for relaxing the cosmological constant
that incorporates some of the features of \Refs{ADKS,KSS}, and that addresses
some of these open questions.
We assume that the standard-model fields are localized on a 3-brane in
6 spacetime dimensions.
We exhibit bulk solutions with the property that no locally-defined quantity
away from the 3-brane is sensitive to the value of the 3-brane tension.
This arises because the only effect of the 3-brane tension on the
geometry is to induce a conical singularity in the transverse space with a
deficit angle proportional to the brane tension.
These solutions have flat induced metric on the 3-brane independent of
the couplings on the 3-brane, and are therefore a natural context for
the ideas of \Refs{ADKS,KSS}.

In order to obtain 4-dimensional gravity at long distances, the extra
dimensions must be compactified.
We first consider several mechanisms for non-singular compactifications that can
be reliably analyzed in the context of 6-dimensional gravity.
We find that none of these mechanisms preserves the property that the
4-dimensional cosmological constant is independent of the couplings
on the 3-brane.
We give general reasons for this.

The solutions that we find generally have naked spacetime singularities at
finite distance from the 3-brane.
These allow new compactification mechanisms that are sensitive to
physics at the 6-dimensional Planck scale.
We speculate that these may allow compactifications with the desired
properties.
We argue that such a compactification cannot be parameterized purely
by higher-dimension operators, suggesting that new Planck-scale degrees of
freedom play an important role.

We then speculate on the behavior of gravity at long distances,
assuming that a compactification mechanism with the desired properties
can be found.
If the compactification scale is $\ell$, we argue that
standard model loop contributions to the effective 4-dimensional cosmological
constant are cut off at distances shorter than $\ell$.
The standard-model contribution to the cosmological constant is then of
order $\ell^{-4} / 16\pi^2$, which can be small if $\ell$ is large.
This model therefore realizes the ideas of \Ref{RamanCosmo}.
Recent cosmological observations suggest that vacuum energy may be a
significant fraction of the present density of the universe,
as suggested by recent observations of type IA supernovae \cite{cosmobs}.
The standard-model contribution to the cosmological constant
is of this order for $\ell \sim 10^{-2}$ mm.
As is now well-known, having 2 extra dimensions of this size 
do not grossly conflict with observation or cosmology \cite{ADD}.
More refined estimates of cosmological bounds
give limits as strong as $\ell \lsim 10^{-5}\mm$ \cite{stronglim}.
We will not attempt to resolve this discrepancy in the present paper, but we
note that the solutions we find have a nontrivial `warp factor' in the bulk that
can weaken the bounds.
It is also interesting that extra dimensions of this size can be probed
in upcoming short-distance gravity experiments and in high-energy collider
experiments.

It is important to remember that the scenario we are discussing is not a
complete solution to the cosmological constant problem.
The reason is that interactions involving bulk fields and their couplings to
other branes must be adjusted to special values to obtain a solution with
unbroken 4-dimensional Poincar\'e invariance.
The reason that the mechanism considered here
is progress is that this tuning does not involve any coupling
involving the 3-brane.
This means that we can hope that the required parameter relations be made
natural by unbroken symmetries in the bulk, for example supersymmetry.

It is interesting that these ideas provide a
completely independent motivation for considering 2 large extra dimensions in
the millimeter range.
The original motivation came from the gauge hierarchy problem \cite{AADD}.
Large extra dimensions can solve the gauge hierarchy problem, and for 2 extra
dimensions the size of the extra dimensions must be of order $1\mm$.
The fact that logarithmic potentials are natural in 2 extra dimensions
gives a natural mechanism for stabilizing large extra dimensions \cite{logstab}.
Finally, precisely in 2 extra dimensions there is a mechanism that may
explain the apparent unification of standard-model couplings near the
Planck scale \cite{dGUT}.

\section{The Mechanism}
To describe the mechanism described above
in more detail, consider a 6-dimensional metric of the
form
\beq[exmetric]
ds^2 = \om^2(r) \eta_{\mu\nu} dx^\mu dx^\nu + dr^2 + \rho^2(r) d\th^2,
\eeq
where $x^\mu$ ($\mu = 0, \ldots, 3$) parameterize the space parallel to
the brane, $\eta_{\mu\nu}$ is the 4-dimensional Minkowski metric,
and $r, \th$ parameterize the transverse space.
This is the most general metric with unbroken 4-dimensional Poincar\'e
invariance and a translation symmetry ($\th \mapsto \th + \hbox{\rm constant}$)
in the transverse space.
We assume that the variable $\th$ is periodic with period $2\pi$, so the
translation symmetry in $\th$ is interpreted as rotation symmetry.
If $\rho(r) \propto r$ as $r \to 0$, then $r = 0$ corresponds
to a single point in the transverse space, like the origin of polar coordinates.
A 3-brane with extent along the $x^\mu$ directions at $r = 0$ induces a conical
singularity with deficit angle
\beq[Detheta]
\De\th = 2\pi \left[ \rho'(0) - 1 \right] = \frac{T_4}{M_6^4},
\eeq
where $T_4$ is the 3-brane tension and $M_6$ is the 6-dimensional
Planck scale.

Away from the brane and away from other singularities, no local quantity is
sensitive to the 3-brane tension.
A simple way to see this is to note that for any value of the brane tension
we can rescale $\th$ so that $\rho'(0) = 1$.
In these coordinates, the deficit angle is parameterized by the range of $\th$,
which cannot be determined by measuring local quantities---as long as
there is no other deficit angle singularities away from $r = 0$.
The reason for the caveat can be understood by representing the solution
with a deficit angle by a smooth metric of the form \Eq{exmetric} with a
`wedge' cut out starting from $r = 0$.
Specifically, we consider two radial geodesics in the transverse space
emerging from the 3-brane with relative angle $\De\th$, and identify
points on the geodesics that are equidistant from the 3-brane.
Away from the brane the position of these geodesics is not measurable, since
their absolute angular position is a coordinate artifact.
However, if the geodesics cross, the space is compactified with spherical
topology, and at the second crossing there is another conical singularity with
a deficit angle proportional to the deficit angle at the original 3-brane.
This must be interpreted as due to a second 3-brane, whose tension is
proportional to the tension of the original 3-brane.
If such singularities are not present, then 
there are solutions for any value of the 3-brane tension.

These solutions must be compactified.
From the discussion above, we see that we do not want to compactify the
solution with a second 3-brane, since that necessarily involves tuning
the two 3-brane tensions against each other.
One promising possibility is to end the space at a $Z_2$ orbifold plane
at constant $r$ surrounding the 3-brane at $r = 0$.
Another possibility is that the space is not compactified, but that the
warp factor $\om(r)$ degreases sufficiently rapidly at large $r$ that
gravity is effectively localized at the 3-brane.

To understand the limitations of these ideas, it is useful to compute
the effective 4-dimensional cosmological constant assuming that a compactified
solution has been obtained.
This can be obtained by integrating the 6-dimensional action evaluated
at the classical solution over the transverse space.
This gives%
\footnote{Our conventions are the same as the book by Wald \cite{Wald}:
the metric is `mostly plus', and $R > 0$ for de Sitter space.}
\beq[Laeffint1]
\La_{4,{\rm eff}} = T_4 -
\myint dr d\th\, \om^4(r) \rho(r) \left[
\frac{M_6^4}{2} R  - \La_6 + \cdots \right],
\eeq
where $R$ is the bulk Ricci scalar, $\La_6$ is the bulk cosmological
constant, and the ellipses denote the contribution from
possible additional bulk fields or other branes.
The Ricci scalar evaluated at the solution can be written
\beq
R = R_{\rm sing} + R_{\rm bulk},
\eeq
where $R_{\rm sing}$ is proportional to $T_4 \de(r)$, and $R_{\rm bulk}$ 
contains the contributions from the bulk fields and other branes.
Precisely for the case of 2 extra dimensions,
the contribution from $R_{\rm sing}$ exactly cancels the contribution
from the first term on the \rhs of \Eq{Laeffint1}, and we have
\beq[Laeffint]
\La_{4,{\rm eff}} = 
-\th_{\rm max} \myint dr\, \om^4(r) \rho(r) \left[
\frac{M_6^4}{2} R_{\rm bulk}  - \La_6 + \cdots \right],
\eeq
where we have performed the (trivial) $\th$ integral.
In order to solve the cosmological constant problem, the integral on the
\rhs of \Eq{Laeffint} must vanish.
This condition is completely independent of couplings on the brane,
which is another manifestation of the independence of the brane tension
discussed above.
However, the vanishing of the integral above is an additional condition
that in general requires fine-tuning \cite{Nilles}.
In the present scenario, we can hope that the vanishing of this integral
can be enforced by unbroken symmetries in the bulk.
For example, supersymmetry can be unbroken in the bulk, and branes away
from the standard-model 3-brane may be BPS states.
This is natural even if supersymmetry is broken badly by fields localized
on the brane, since radiative corrections of brane fields do not renormalize
the couplings associated with bulk modes and distant branes.

\section{Bulk Solutions}
We begin by determining the metric away from the branes.
The solution is a special case of the one obtained in \Ref{CP}
(see also \cite{RS}), but we
will derive it from first principles for completeness.
We generalize the metric to allow constant curvature on the 3-brane.
We parameterize the 6-dimensional metric as
\beq[genmetric]
ds^2 = \om^2 \ga_{\mu\nu} dx^\mu dx^\nu
+ s(\om) d\om^2 + f(\om) d\xi^2,
\eeq
where $\ga_{\mu\nu}$ is a metric with constant curvature $\la$
(\ie $R_{\mu\nu}(\ga) = \la \ga_{\mu\nu}$).
The `warp factor' $\om$ is the radial coordinate in the transverse
space, while $\xi$ is an angular variable.
The solution is particularly simple in these coordinates, as we will see.
(These coordinates are also employed in \Ref{CP}.)
The nonvanishing components of the Einstein tensor are
\beq\bal
G_{\mu\nu} &= \frac{1}{s} \left[
\frac{f''}{2 f} - \frac{f'^2}{4 f^2} - \frac{f' s'}{4 f s}
+ \frac{3 f'}{2 \om f} - \frac{3 s'}{2 \om s} + \frac{3}{\om^2}
- \frac{\la s}{\om^2}
\right] g_{\mu\nu},
\\
G_{\om\om} &= \frac{1}{s} \left[
\frac{2 f'}{\om f} + \frac{6}{\om^2} - \frac{2\la s}{\om^2} \right]
g_{\om\om},
\\
G_{\xi\xi} &= \frac{1}{s} \left[
-\frac{2 s'}{\om s} + \frac{6}{\om^2} - \frac{2\la s}{\om^2}
\right] g_{\xi\xi},
\eal\eeq
where $f' = df / d\om$, \etc\ \ 

The bulk cosmological constant and the 4-dimensional curvature $\la$
cancel in the difference of the $\om\om$ and $\xi\xi$ equations,
which give $f s' + f' s = 0$.
This immediately implies
\beq[seq]
s(\om) = \frac{1}{f(\om)},
\eeq
where a possible overall constant can be set to one by rescaling
$x^\mu$ and $\om$.
Substituting this into the $\xi\xi$ or $\om\om$ equations gives
\beq
\om f' + 3 f = -\frac{\om^2}{2} \frac{\La_6}{M_6^4} + \la.
\eeq
The general solution is simply
\beq[feq]
f(\om) = -k \om^2 + \sfrac{1}{3}\la + c \om^{-3},
\eeq
where
\beq
k = \frac{\La_6}{10 M_6^4}
\eeq
and $c$ is an arbitrary integration constant.
With this solution for $f(\om)$ and $s(\om)$, one finds that the $\mu\nu$
equations are solved;
this is a general consequence of the reparameterization invariance in $r$.
Note that $\om$ is dimensionless, $k$, $\la$, and $c$ have mass dimension $+2$,
and $\xi$ has mass dimension $-2$.

The relation to the more `physical' coordinates of \Eq{exmetric}
is easy to work out from
\beq
dr = \pm \frac{d\om}{\sqrt{f(\om)}},
\eeq
where the sign indicates whether the warp factor
is increasing or decreasing with increasing $r$.
For $\la = 0$, we obtain
\beq[omr]
\om(r) &= \begin{cases}
\left[ \cos \al (r - r_0) \right]^{2/5}
& for $\La_6 > 0$,
\\
\left[ \cosh \al (r - r_0) \right]^{2/5}
& for $\La_6 < 0$,
\end{cases}
\\
\eql{rhoconst}
\rho(r) &= \rho_0 \om'(r),
\eeq
where
\beq
\al \equiv \sqrt{\frac{5 |\La_6|}{8 M_6^4}}.
\eeq

We now discuss the physical interpretation of the solutions for $\la = 0$.
In the coordinates of \Eq{genmetric},
the physical region corresponds to $f(\om) > 0$.
For $c = 0$, we must have $k < 0$, and the only solution is anti de Sitter space.
For $c \ne 0$, we can choose $c = k$ by rescaling the coordinates, so that the
physical region corresponds to $0 \le \om \le 1$ for $k > 0$, and $\om \ge 1$
or $\om < 0$ for for $k < 0$.
For both signs of $k$, $f$ has a simple zero at $\om = 1$.
The geometry there can be exhibited by transforming to the $r, \th$
coordinates:
\beq[rorg]
r - r_0 
\simeq \frac{2}{\sqrt{5 |k|}} \sqrt{|\om - 1|}
\quad \hbox{\rm for}\ \om \simeq 1.
\eeq
The metric near $\om = 1$ is then
\beq
ds^2 \simeq \eta_{\mu\nu} dx^\mu dx^\nu + dr^2 + r^2 d\theta^2,
\qquad
\th \equiv \frac{5|k|}{2} \xi,
\eeq
where we have chosen the constant $\rho_0$ in \Eq{rhoconst} to fix
the coefficient of $d\th^2$.
If $\th$ is periodic with period $\th_{\rm max} = 2\pi$, there is no singularity
at $\om = 1$ ($r = 0$).
For any other value of $\th_{\rm max}$ there is a deficit angle singularity at
$r = 0$ that we interpret as being due to the presence of a 3-brane at $r = 0$
with tension proportional to the deficit angle.%
\footnote{We have verified this explicitly by smearing out the 3-brane by
replacing it by a `ring' at $r = a$ and solving the gravitational field
equations.
We find that the deficit angle is the only effect that survives in the
limit $a \to 0$ with the 3-brane tension held fixed.}
The 3-brane equations of motion impose the condition $\om'(r_0) = 0$.
We can see that this is satisfied from \Eq{omr}.

For $k < 0$, the solution given by \Eqs{seq} and \eq{feq} for $\om \ge 1$
describes a 3-brane in infinite space of negative
cosmological constant.
The `warp factor' $\om$ increases away from the 3-brane, so gravity is not
localized.

For $k > 0$, $\om$ decreases away from the 3-brane,
and $f(\om)$ diverges as $\om \to 0$.
This is a true singularity, as can be seen from the curvature invariant
\beq
R_{MNPQ} R^{MNPQ} \simeq \frac{240 k^2}{\om^{10}}
\quad \hbox{\rm as}\ \om \to 0,
\eeq
where $M, N = 0, \ldots, 5$.
The proper distance to the singularity is $\pi / (5\sqrt{k})$, and
the time measured by an observer on the brane for a light signal to go from the
brane to the singularity and back is
\beq
\De t = 2 \int_0^1 d\om\, \frac{1}{\om \sqrt{f(\om)}}
\simeq \frac{1.8}{\sqrt{k}}.
\eeq
This singularity is similar to the one found in \Ref{CK}.
It is an interesting conjecture (following \Refs{ADKS,KSS}) that in a more
fundamental theory of gravity this singularity is smoothed out in a way that
preserves the independence of the solution on the 3-brane tension.

For $k < 0$ the region $\om < 0$ is also physical, although it is not
possible to have a 3-brane in this region.
(Note that the metric depends only on $\om^2$, so the sign of $\om$ is
not physical.)
This also has a naked singularity as $\om \to 0$;
in fact, the metric is the same near both singularities.

\section{Compactification}
We now turn to compactification.
We first attempt to construct compactified solutions using mechanisms that
we can control in the 6-dimensional gravity effective theory.
We will find that these conventional mechanisms fail for very general
reasons.
We then turn to possible compactifications that depend on the existence
of the naked singularities.

\subsection{Minimal 4-Branes}
We first attempt to find a compact solution with a 4-brane surrounding
the 3-brane at fixed $r$.%
\footnote{Solutions similar to the one constructed here are considered in \Ref{CN}.}
In order to avoid a second deficit angle singularity, we assume that the
4-brane is at a $Z_2$ orbifold boundary, which effectively ends the space.

It is easy to see that there are no solutions if the 4-brane action contains
only a tension term.
The reason is that the 4-brane tension gives rise to boundary conditions
(in the metric of \Eq{exmetric})
\beq
\frac{\De\om'}{\om} = \frac{\De\rho'}{\rho} = -\frac{T_5}{4}.
\eeq
At a symmetric point, this implies that
$\om'/\om = \rho'/\rho$ on both sides of the 4-brane.
If we use this as initial data for evolution in $r$, we obtain a solution
with $\rho(r) \propto \om(r)$.
This cannot evolve in a finite distance to the
solution near the 3-brane, which has
$\rho(r) \not\propto \om(r)$.

\subsection{Non-minimal 4-Branes}
We now consider the possibility that there is a non-minimal
stress tensor on the 4-brane of the form
\beq[Ttheta]
t_{\mu\nu} = -T_5 \ga_{\mu\nu},
\quad
t_{\th\th} = -T_{5,\th} \ga_{\th\th},
\eeq
where $\ga$ is the induced metric on the 4-brane.
This is the most general stress tensor compatible with the symmetries we
are assuming.
The asymmetry $T_{5,\th} \ne T_5$ allows us to find solutions with a
3-brane surrounded by a 4-brane at finite distance, as we will see below.
In the limit $T_{5,\th} \to T_5$ the proper separation between the 3-brane
and the 4-brane
becomes infinite, so the mechanism that gives rise to the asymmetry is
crucial for the compactification.

To cancel the standard-model contributions to the cosmological
constant, it is important that the mechanism that gives rise to the asymmetry
does not bring in dependence on the 3-brane tension via the deficit angle.%
\footnote{We thank A. Nelson and R. Sundrum for emphasizing this point.}
For example, Casimir energy from fields localized on the 4-brane will
give rise to a stress tensor of the form \Eq{Ttheta}, but
$T_{5,\th} - T_5$ is proportional to the circumference of the 4-brane,
which depends on the deficit angle at the 3-brane.
Similarly, a scalar field localized on the 4-brane that wraps in the $\th$
direction must come back to its original value around the 4-brane, and
hence its stress tensor also depends on the deficit angle.

A more promising mechanism is to localize a 3-form gauge field on the 4-brane
and assume that it has nonzero field strength
\beq
F_{\mu\nu\rho\si} = E \ep_{\mu\nu\rho\si},
\qquad E = \hbox{\rm constant},
\eeq
with all other components vanishing.
This is a constant `electric' field in the $x^\mu$ directions, and is not
subject to any quantization condition.
The sign of the stress tensor from this configuration gives
$T_5 - T_{5,\th} > 0$.
We will see that this is the sign that we require to obtain a solution.
The solution requires a single fine-tuning involving $E$, but this fine-tuning
does \emph{not} involve the 3-brane tension.

Nonetheless, this is not a satisfactory solution.
To understand the reason, consider adiabatically changing the 3-brane tension.
This can be thought of as a crude model for a phase transition involving
matter on the 3-brane.
The metric then has the form \Eq{exmetric} with an adiabatically
evolving deficit angle.
As the deficit angle changes in time, the size of the 4-brane changes.
This changes the value of $E$, since it is the integrated number of field lines
(rather than $E$) that is conserved in the adiabatic evolution.
The 4-form flux mechanism has effectively recast the cosmological constant
problem as a problem of fine-tuning initial data.

This phenomenon is actually a general consequence of the conservation of
stress-energy.
To see this, consider a stress tensor of the form \Eq{Ttheta} in a metric
with adiabatically changing deficit angle.
This metric can be parameterized by the replacement $\rho(r) \to \rho(r, t)$
in \Eq{exmetric} (with a fixed range for $\th$).
The equation $\nabla^M T_{M 0} = 0$ then implies
\beq
\partial_t T_5 =
\left( T_{5,\theta} - T_5 \right) \frac{\partial_t \rho}{\rho}.
\eeq
We see that $T_5$ must change in response to the deficit angle.

In the remainder of this section, we construct the solutions explicitly.
The main goal of this is obtain a quantitative estimate of the fine-tuning of
the 4-brane tension that is required to obtain a small cosmological constant.
We find that there is no improvement relative to a generic theory with new
physics at the TeV scale.
The reader who is not interested in these details can skip
to the next Subsection.

Away from the 4-brane, the solution has the form considered in the previous
Section.
At the 4-brane, the metric components have discontinuous first derivatives.
In terms of the metric \Eq{exmetric},
the relevant terms in the Einstein tensor are those with second derivatives
with respect to $r$.
These are
\beq[match]\bal
G_{\mu\nu} &= \left[
\frac{3 \om''}{\om} + \frac{\rho''}{\rho} + \cdots
\right] g_{\mu\nu},
\\
G_{\th\th} &= \left[
\frac{4 \om''}{\om} + \cdots
\right] g_{\th\th}.
\eal\eeq
The 4-brane gives a nonvanishing contribution to the $\mu\nu$ and $\th\th$
components of the stress tensor proportional to $\de(r - r_0)$,
where $r_0$ is the radial position of the 4-brane.
This gives the discontinuity conditions
\beq
\frac{\De \om'}{\om} = -\frac{T_{5,\th}}{4},
\qquad
\frac{\De \rho'}{\rho} = -T_5 +\frac{3 T_{5,\th}}{4},
\eeq
where 
we now use units where $M_6 = 1$.
The functions $\om$ and $\rho$ are continuous at the 4-brane.

It is convenient to write these conditions in terms of the `warp factor'
coordinates using
\beq
\om' = \ep \sqrt{f(\om)},
\qquad \ep = \pm 1.
\eeq
The sign factor $\ep$ tells us whether the warp factor is increasing or
decreasing with increasing $r$;
it can be chosen independently on either side of the brane.
We choose the coordinate $r$ to increase  monotonically as we pass through
the 4-brane from the 3-brane position at $r = 0$.

Because of the $Z_2$ orbifold projection, we require the solution to be
symmetric with respect to reflections about the 4-brane.
This means that the bulk solution is described by the same function
$f(\om)$ on both sides of the 4-brane.
The $\om'$ discontinuity equation is then
\beq[omjumpf]
\frac{\sqrt{f_0}}{\om_0} \left( \ep_2  - \ep_1 \right)
= -\frac{T_{5,\th}}{4},
\qquad
f_0 \equiv f(\om_0),
\eeq
where $\om_0$ is the warp factor at the 4-brane.

We first consider a space with negative cosmological constant.
The warp factor increases away from the 3-brane, so $\ep_1 = +1$,
$\ep_2 = -1$.
From \Eq{omjumpf} we see that this requires $T_{5,\th} > 0$.
\Eq{omjumpf} can then be written as
\beq[omeq]
-k + \sfrac{1}{3} \la \om_0^{-2} + c \om_0^{-5} = \left( \frac{T_{5,\th}}{8} \right)^2.
\eeq
Using
\beq
\frac{\rho'}{\rho} = \ep \frac{f'(\om)}{2\sqrt{f(\om)}},
\eeq
the $\rho'$ discontinuity equation can be written
\beq
-\frac{f'(\om_0)}{\sqrt{f_0}} = -T_5 +\frac{3 T_{5,\th}}{4}.
\eeq
Combined with \Eq{omeq}, this can be simplified to give
\beq[rhoeq]
-2 k - 3 c \om_0^{-5} = \frac{T_{5,\th}}{8}
\left( T_5 - \sfrac{3}{4} T_{5,\th} \right).
\eeq
For $\la = 0$, we can combine this with \Eq{omeq} to obtain
\beq[ftune]
k = -\frac{T_{5,\th}}{40} \left( T_5 - \sfrac{3}{8} T_{5,\th} \right).
\eeq
This is a fine-tuning condition that is required to obtain a solution with
vanishing 4-dimensional curvature.
For $\la \ne 0$, \Eqs{omeq} and \eq{rhoeq} can be solved for $\om_0$ and
$\la$.
A similar fine-tuning in \Ref{RS1} is required to ensure the flatness
of a brane with codimension 1.
In fact, if we take the limit $c \to 0$, $T_{5,\th} \to T_5$,
we obtain exactly the solution of \Ref{RS1} in one dimension higher.

We now assume that $c \ne 0$, so that we can rescale coordinates and choose
$c = k$.
We also set $\la = 0$.
The solution then has a deficit angle at $\om = 1$
that we interpret as a 3-brane.
The 4-brane position $\om = \om_0$ is then determined by \Eqs{omeq}
and \Eq{ftune} to be
\beq[warp4]
\om_0^{-5} = \frac{T_{5,\th}}{40 k}\left(T_{5,\th} - T_5 \right).
\eeq
The proper radial distance between the 3-brane and 4-brane is
\beq
\ell = \frac{1}{\sqrt{-k}} \int_1^{\om_0} \frac{d\om}{\sqrt{\om^2 - \om^{-3}}}
= \frac{2}{5 \sqrt{|k|}} \ln \left[ \sqrt{\om_0^5} + \sqrt{\om_0^5 - 1} \right].
\eeq
As discussed above, in the limit
$T_{5,\th} \to T_5$, we have $\om_0 \to \infty$ and $\ell \to \infty$.
In order to obtain finite $\ell$ with $\om_0 > 1$, we need $T_5 > T_{5,\th}$
(remember $k < 0$).
These signs are compatible with the stress tensor from a 3-form gauge field,
as discussed previously.
The effective 4-dimensional Planck scale is given by
\beq[effm4]
M_4^2 = \frac{2 M_6^4}{15 |k|}  \left( \om_0^3 - 1 \right) \th_{\rm max}.
\eeq
The warp factor has maximum value at the 4-brane, so the solution
tends to localize gravity there.
The warp factor grows exponentially as a function of $\sqrt{-k}\,\ell$
(see \Eq{omr}), so we can solve the gauge hierarchy problem via the
mechanism of \Ref{RS2} for $\ell \sim 10 / \sqrt{-k}$.
We will not pursue this possibility here.

For $k > 0$, a similar analysis shows that \Eqs{ftune} and \eq{warp4}
hold in this case as well, but now $T_{5,\th} < 0$ and $T_5 < T_{5,\th}$.
The fact that the 4-brane has negative tension is not a problem, since
it is fixed on the orbifold plane.
In these solutions, the warp factor decreases away from the 3-brane,
and the 4-brane cuts of the space before the naked singularity at $\om = 0$.

We now consider the fine-tuning of the cosmological constant in these
solutions.
If we perturb the 4-brane tensions away from their fine-tuned values,
$T_5 \to T_5 + \De T_5$, $T_{5,\th} \to T_{5,\th} + \De T_{5,\th}$,
we obtain
\beq
\la &= \frac{\om_0^2}{8} \left[
T_{5,\th} \De T_5 + \left(T_5 - \sfrac{3}{4} T_{5,\th} \right) \De T_{5,\th}
\right] + \scr{O}(\De T^2),
\\
\De\om_0 &= -\frac{\om_0^6}{120 k} \left[
T_{5,\th} \De T_5 + \left(T_5 - \sfrac{3}{2} T_{5,\th} \right) \De T_{5,\th}
\right] + \scr{O}(\De T^2).
\eeq
We estimate the 4-dimensional curvature $\la$ for an order-1 perturbation
of the 3-brane tension $\De T_4 \sim T_4$.
This results in a change of the deficit angle $\De\th \sim \De T_4 / M_6^4$,
which may be small if $M_6 \gg T_4$.
(This can be natural due to approximate SUSY on the 3-brane.)
Also, $\De T_5$ and $\De T_{5,\th}$ are naturally of order
$(T_5 - T_{5,\th}) \De\th$, where we allow $T_5 - T_{5,\th} \ll T_5$.
This results in a 4-dimensional curvature
\beq
\la \sim \frac{\om_0^2 T_5^2 \De T_4}{M_6^{4}}.
\eeq
Combining this with \Eqs{ftune}, \eq{warp4}, and \eq{effm4}, we obtain
\beq
\la \gsim \frac{\De T_4}{M_4^2}.
\eeq
Since $\De T_4 \gsim 1\TeV$, this is no better than a 4-dimensional
theory with a TeV scale vacuum energy (\eg a theory with supersymmetry
broken at the TeV scale).

\subsection{Warp Factor Compactification}
We now consider the possibility that the extra dimensions are infinite
but with a `warp factor' $\om(r)$ that decreases sufficiently rapidly
as $r \to \infty$ so that gravity is approximately 4-dimensional at
long distances \cite{RS2}.
In this scenario, since gravity is not 4-dimensional even at arbitrarily long
distances, it is obvious how it evades `no-go' theorems concerning the tuning of
the cosmological constant \cite{Weinberg}.

The simplest possibility would be to surround the 3-brane by a 4-brane
in the extra dimensions, with the spacetime being anti de Sitter on the
outside of the 4-brane.%
\footnote{In principle the solution need only approach anti de Sitter at
large distances, but the only solution with the required symmetries is
exactly  anti de Sitter.}
However, the anti de Sitter metric has the symmetry $\rho(r) \propto \om(r)$,
which is preserved by the brane matching condition.
Therefore, anti de Sitter space cannot match onto the asymmetric solution
that is required if we have a 3-brane in the solution.

Another possibility is to have additional fields in the bulk, and look for
solutions where the warp factor vanishes at infinity.
This can evade the obstruction described above, since an asymmetric solution
with $\rho(r) \not\propto \om(r)$ can evolve to a symmetric solution
asymptotically.
The simplest possibility is to introduce scalars $\phi_a$ into the bulk:
\beq
S_{\rm scalar} = \myint d^6 x\, \sqrt{-g} \left[
-\frac{1}{2} g^{MN} \partial_M \phi_a \partial_N \phi_a - V(\phi) \right].
\eeq
The resulting Einstein equations can be simplified by introducing the
quantities
\beq
\Si(r) \df \frac{1}{\sqrt{-g}} \partial_r \sqrt{-g}
= \frac{4 \om'}{\om} + \frac{\rho'}{\rho},
\qquad
\De(r) \df \frac{\om'}{\om} - \frac{\rho'}{\rho}.
\eeq
We require $\sqrt{-g}$ to decrease as $r \to \infty$ to have finite
volume, so $\Si < 0$.
$\De$ is a measure of the asymmetry between the $\th$ and $x^\mu$ directions.
In terms of $\Si$ and $\De$, Einstein's equations are first-order
differential equations.
Taking the difference of the $\mu\nu$ and $\th\th$ components of Einstein's
equations gives
\beq[simpeq1]
\frac{\De'}{\De} = -\Si.
\eeq
Since $\Si < 0$, we see that $|\De|$ increases monotonically
as $r \to \infty$.
This means that we cannot hope to obtain a solution that approaches
anti de Sitter space (which has $\rho(r) \propto \om(r)$ and hence
$\De \equiv 0$) with a decreasing warp factor.

We may still hope to find compactified solutions with a decreasing warp
factor that are not anti de Sitter at infinity.
Using \Eq{simpeq1} to simplify the difference of the $\th\th$ and $rr$ Einstein
equations gives
\beq[simpeq2]
\Si' = -\De^2 - \sfrac{5}{4} \phi'_a \phi'_a.
\eeq
Because the \rhs is negative-definite, $\Si$ decreases (becomes more negative)
monotonically.
It can be shown that all solutions with a decreasing warp factor have
a singularity at finite $r$ by using the fact that the solutions are bounded
by the solutions with $\phi'_a \equiv 0$.
In detail, let $\Si(r)$ and $\De(r)$ be solutions with some initial conditions
at $r = r_0$, and let $\Si_0(r)$ and $\De_0(r)$ be the solutions of
\Eqs{simpeq1} and \eq{simpeq2} with the
same initial conditions, but setting $\phi'_a \equiv 0$.
We then have $\Si(r) < \Si_0(r)$ and hence $|\De(r)| > |\De_0(r)|$ for
$r > r_0$.
It is not hard to see that $|\De_0(r)| \to \infty$ at finite $r$, so
$\De(r)$ must also have a singularity at finite $r$.

\subsection{Singular Compactification?}
A striking feature of the solutions constructed in the previous Section
is the presence of naked singularities.
Near a naked singularity the curvature is blowing up, so the physics of
the singularity is sensitive to the details of physics above the Planck
scale.
It is natural to speculate (following \Refs{ADKS,KSS})
that this new physics ends the space at the
singularity.
In this scenario, the long-distance behavior of gravity is directly controlled
by the short-distance behavior at the singularity,
and we cannot rigorously address the physics of the singularity using the low-energy
effective theory.
We will therefore confine ourselves to some simple observations.

First, we note that the curvature distinguishes between the
$x^\mu$ and $\th$ directions.
(For example, $R_{r \mu r \mu} \ne R_{r \th r \th}$ for $\mu = 0, \ldots, 3$.)
This is potentially important for compactification because the failure of
the compactification mechanisms above can be traced to the
non-existence of terms in the equations of motion that distinguish between the
$x^\mu$ and $\th$ directions independently of the 3-brane tension.
Near the naked singularity, these higher-derivative effects may be important
and allow more general boundary conditions than the ones we considered.
The resulting theory would be a $Z_2$ orbifold with Planck-scale
curvature near the orbifold boundary.

We can attempt to get some insight into this scenario
by adding higher-derivative terms to the action.
We cannot hope to obtain consistent solutions, but we can see that
the hoped-for effects do not occur if we treat the higher-derivative
terms perturbatively.
At each order in perturbation theory, we have a system of second-order equations
that involves the lower-order solutions.
At each order, the boundary conditions on the metric at the
orbifold boundary is $\om'/\om = \rho'/\rho = 0$, which gives
rise to a symmetric solution ($\rho \propto \om$).
For higher-derivative terms localized on the orbifold boundary, the
same argument holds if we regulate the brane (\eg by a scalar domain wall).

If we attempt to include the higher-derivative terms non-perturbatively,
the initial data at the orbifold boundary involves higher derivatives of
the metric.
Taking asymmetric initial conditions, one can presumably find solutions
of the kind we seek.
However, it is difficult to interpret such solutions physically.
The additional initial conditions can be thought of as additional
Planck-scale degrees of freedom.
However, theories with higher-derivative terms are generally classically
unstable, corresponding to the fact that the extra degrees of freedom are
generally ghosts.
At best, this line of reasoning may be viewed as a weak hint
that new Planck-scale degrees of freedom localized at the orbifold
boundary may allow compactification.

Recently, there has been an interesting proposal to make sense out of
naked singularities such as this by imposing boundary conditions at
the singularity \cite{CK}.
While we regard this proposal as very interesting, we note that it appears
to be difficult to give a generally covariant formulation of the
boundary conditions for the metric (see however \Ref{Waldsing}).
Another interesting approach is to look for solutions that
`regulate' the singularity by hiding it behind an event horizon \cite{Gubser}.
These approaches are definitely worthy of further exploration.

We conclude that there are potential difficulties with the idea of
compactification near the naked singularities, but the idea cannot be
ruled out.
In light of this, we believe that the scenario described here is
worth further investigation.

\section{Effective Field Theory}
We now turn to the effective field theory analysis of the scenario
described above, assuming that compactification is possible.
Without a specific compactification mechanism, we cannot address the
details of the Kaluza-Klein (KK) spectrum or the 4-dimensional effective
field theory, but we can analyze some simple aspects of the scenario
that depend on the behavior near the 3-brane.

We first consider the light degrees of freedom.
The KK spectrum contains the 4-dimensional graviton and a massless gauge field
corresponding to the unbroken rotational invariance in the transverse space.
These can be parameterized by the metric
\beq
ds^2 = \om^2(r) g_{\mu\nu}(x) dx^\mu dx^\nu
+ dr^2 + \rho^2(r) \left[ d\th + A_{\mu}(x) dx^\mu \right]^2,
\eeq
where $\om(r)$ and $\rho(r)$ are as in our solution.
The KK gauge field couples to momentum in the $\th$ direction,
so the standard-model fields are not `charged' under the gauge group.
In addition, the couplings of the KK gauge field is suppressed by large
volume factors and the wavefunction factor $\rho(r)$ above.
The bounds on the couplings of such a vector are therefore much weaker
than the corresponding bounds on KK gravitons, which is safe.

Another important question we can address is whether higher-order terms
on the brane upset the non-dependence of the 4-dimensional cosmological
constant on the 3-brane tension.
For example, we can write
\beq[ct]
\De S_{\rm brane} \sim \myint d^4 x \sqrt{-\ga}\, M_6^2 \left[
g^{MN} R_{MN} + \ga^{\mu\nu} R_{\mu\nu} + \cdots \right].
\eeq
Such terms can be generated by standard-model loops with external
gravitational lines;
in fact, \emph{all} such loop effects proportional to positive powers of
$M_6$ (the cutoff) correspond to ultraviolet divergences, and therefore
correspond to operators localized on the brane.
Since all such effects are equivalent to tree-level terms,
this reduces the question to a classical analysis.

Making sense of the equations of motion that follow from \Eq{ct}
requires regulating the 3-brane, presumably taking care to preserve
general covariance.
However, a very general argument shows that the independence of the
4-dimensional cosmological constant of the 3-brane tension is robust against the
addition of such effects.
Let us momentarily adopt the contrary hypothesis, namely that the
absence (or fine-tuning) of the couplings in \Eq{ct} are required to obtain a
solution with unbroken 4-dimensional Poincar\'e invariance.
We should then ask what the solutions are.
If we violate the fine-tuning of the couplings away from the 3-brane by a small
amount, it is clear that the effect in the 4-dimensional effective field theory
is a nonzero cosmological constant.
The vacuum solutions in the 4-dimensional field theory are then de Sitter or
anti de Sitter space.
The 6-dimensional metric that corresponds to this solution should therefore
have the symmetries of 4-dimensional de Sitter or anti de Sitter space;
this metric was given in \Eqs{genmetric}, \eq{seq}, and \eq{feq}, where
the parameter $\la$ is the 4-dimensional curvature.
This metric admits deficit angle singularities at simple zeros of
$f(\om)$.
Since these are the \emph{only} solutions with the required symmetries,
we conclude that all the effects of the 3-brane couplings can be absorbed
into the deficit angle and the 4-dimensional curvature.
(This is analogous to a `no-hair' theorem for codimension 2 branes.)

We must still address the possibility that the 4-dimensional curvature is
sensitive to the terms in \Eq{ct}.
When these terms are properly regulated, their effect on physics
below the scale $M_6$ can be written as boundary conditions on the
gravitational fields involving higher derivatives.
If the terms in \Eq{ct} can be treated as non-singular perturbations,
these new boundary conditions will still allow solutions for any
value of the 4-dimensional curvature, including flat space.
Under our assumptions, it is the matching condition at the \emph{other}
boundary of the transverse space that picks out the value of $\la$.
We conclude that the presence of the terms \Eq{ct} does not invalidate
our picture.

This argument eliminates contributions to the effective 4-dimensional cosmological
constant proportional to positive powers of $M_6$.
However, we expect loop matching corrections from loops of
standard-model fields of order
\beq[Laeff]
\La_{4,{\rm eff}} \sim \frac{\ell^{-4}}{16\pi^2}.
\eeq
The scale of these corrections is set by $\ell$ because the size of the
dominant loops in position space is of order $\ell$.
There are no contributions involving positive powers of $M_6$ because these
would have to correspond to a local counterterm in the 6-dimensional
theory that gives an $\ell$-dependent contribution to $\La_{4,{\rm eff}}$
at tree level.%
\footnote{%
In a Kaluza-Klein description, the 6-dimensional bulk fields are rewritten as an
infinite tower of 4-dimensional fields. 
One might worry that a single KK state will give a contribution of
order $m_{\rm KK}^4$ to the effective 4-dimensional cosmological constant,
which will be much larger than \Eq{Laeff} for large $m_{\rm KK}$.
However, the different KK states are nothing more than different
eigenstates of momentum in the compact directions, and so such contributions
correspond to contributions proportional to the cutoff $M_6$ from loops
with high momentum.
The argument above shows that these cannot occur.
The underlying reason is 6-dimensional locality, which is not manifest in
a KK description.}

We can ask whether it is possible that $\La_{4,{\rm eff}}$ could be
\emph{smaller} than the estimate \Eq{Laeff}.
We believe that this is impossible, simply because of the naturalness of the
effective theory at distances larger than $\ell$.
There are no massless scalars in this effective theory, and hence no light
degrees of freedom that can adjust the cosmological constant to zero.

We can get some insight into the mechanism for the cancellation of the
standard model contribution to the cosmological constant by considering
the dynamics of a slowly rolling scalar field localized on the 3-brane.
The potential of the scalar acts as an effective 3-brane tension that varies
with time.
If the rate of change of this tension is sufficiently slow, the bulk
gravitation fields will respond adiabatically.
The solution will therefore have a deficit angle that tracks the instantaneous
value of the scalar potential.
From the point of view of the 4-dimensional effective theory, this corresponds
to a mixing between the scalar and the gravity KK mode that corresponds to the
deficit angle.
This light mode adjusts itself to cancel the cosmological constant.
We expect this mode to have a mass of order $1\mm^{-1}$, and below this scale
the cancellation mechanism is no longer effective;
this leads to the estimate \Eq{Laeff}.
However, we do not understand the generation of a nonzero cosmological
constant from a 6-dimensional perspective.

The arguments above address only the contribution of standard-model loops to
the effective cosmological constant.
We emphasize again that even if one accepts the existence of a compactification
with the properties described in the previous Section, we do not have a
complete solution to the cosmological constant problem.
In particular, we have not addressed the question of bulk gravity loops,
and we have seen that fine-tuning of bulk interactions
is necessary to obtain a solution with unbroken 4-dimensional Poincar\'e
invariance.
The important point is that the quantities that must be fine-tuned do not
involve the 3-brane couplings.
We can therefore hope that unbroken bulk symmetries such as supersymmetry
can make these parameter choices natural.
We therefore believe that it is plausible that the leading contribution to the
cosmological constant is of order \Eq{Laeff}.

\section{Conclusions}
We have described a natural mechanism for canceling the standard-model
contribution to the cosmological constant.
It relies only on the properties of branes with 2 transverse dimensions.
The mechanism requires the compactification scale to be in the millimeter
range, and suggests a nonzero cosmological constant in the range favored
by cosmology.
Previous work has shown that the presence of 2 large extra dimensions can also
explain the gauge hierarchy problem and the unification of gauge couplings;
this confluence of ideas is nothing if not suggestive.
Most importantly, these ideas are testable by terrestrial experiments and
cosmological observations.

While we have not definitely established all aspects of the mechanism we
proposed, we hope that some of these ideas will prove fruitful in the
search for the ultimate solution of the cosmological constant problem.


\section*{Acknowledgments}
We thank A. Nelson and R. Sundrum for important comments on an earlier
version of this paper.
We also thank N. Arkani-Hamed, T. Jacobson, R. Sundrum, and J. Terning
for discussions.
This work was supported by the National Science Foundation under
grant PHY-98-02551, and by the Department of Energy under grant
DOE/ER/40762-204.

\end{document}